# The Extended Edit Distance Metric

Muhammad Marwan Muhammad Fuad and Pierre-François Marteau

VALORIA, Université de Bretagne Sud
BP 573, 56017 Vannes
{marwan.fuad, Pierre-francois.marteau}@univ-ubs.fr

**Abstract.** Similarity search is an important problem in information retrieval. This similarity is based on a distance. Symbolic representation of time series has attracted many researchers recently, since it reduces the dimensionality of these high dimensional data objects. We propose a new distance metric that is applied to symbolic data objects and we test it on time series data bases in a classification task. We compare it to other distances that are well known in the literature for symbolic data objects. We also prove, mathematically, that our distance is metric.

**Keywords:** Time Series, Symbolic Representation, SAX , Edit Distance, Extended Edit Distance.

## 1 Introduction

The problem of similarity search and information retrieval in large databases has received an increasing deal of attention recently, because of its vast number of applications. This problem, because of its nature, is not a trivial one, since in most cases the databases in question are very large. So using sequential scanning to search data can take a long time and can become ineffective, especially when the data objects stored in modern databases are complex. For all these reasons, there is always a need to introduce new methods to deal with this problem.

Research in this area has focused on different characteristics. One of them is the distance metric that is used to support the similarity search mechanism.

Many distance metrics have been suggested. But the Euclidean distance is still the most widely used, even though it has many drawbacks.

Another characteristic of the search problem is data representation. In multimedia search the main problem we encounter is the so called "dimensionality curse". The main technique used to deal with this dimensionality curse, which is, in fact, the essence of data representation, is data compression.

In time series, for example, there have been different suggestions to represent them. To mention a few; DFT [1] and [2], DWT [3], SVD [11],APCA [10],PAA [9] and [16], PLA [12], SAX [7], …etc.

Among data compression techniques, symbolic representation is an idea that seemed to have potentially interesting pros, in that by using it we can benefit from the wealth of text-retrieval algorithms and techniques. However, the first papers presented were mainly ad hoc. In addition, they didn't present a technique to support Euclidean queries. There were also other questions concerning the discretization and the size of the alphabet [10].

But symbolic representation is receiving more and more attention. New distance measures mainly adapted to this kind of representation have been proposed. Also there have been many papers that suggest methods to discretize the data. For all these reasons, symbolic representation seems very promising.

## 2 Motivations

### 2.1 The Edit Distance

The edit distance (ED) is defined as the minimum number of delete, insert, and substitute operations needed to transform string S into string T. This distance is the main distance measure used to compare two strings is the edit distance (ED). This distance uses three operations: insert, delete, and change [15]. Different variations of this distance were proposed later like the edit distance on real sequence (EDR) [4], and the edit distance with real penalty (EDRP) [4]

The edit distance has a main drawback, in that it penalizes all change operations in the same way, without taking into account the character that is used in the change operation. This drawback is due to the fact that the edit distance is based on local procedures, both in the way it's defined and in the algorithms used to compute it. This poses questions on the accuracy of the similarities obtained by applying this distance. We will give herewith two examples to show this;

**Example 1:**

The edit distance was presented mainly to apply on spelling errors. But because of the conventional keyboard arrangement, the probability that an "A" be mistyped as "S" is not the same as mistyping "A" as "P", for instance (on an English keyboard), but yet, the edit distance doesn't take these different possibilities into consideration.

**Example 2:**

In the In time series databases, some methods of dimensionality reduction, like the one presented by Keogh [7], use symbolic representation. These methods are based on converting the time series into an alphabet of discrete symbols. Then the textual processing techniques can be used to manage these strings. These symbolic representation methods, as much as we know, don't take into consideration which character was changed by which character.

One of the ways that can be considered to deal with this problem is to use a predefined table that shows the cost of change between any two characters of that alphabet. This method, although worth considering, has a few cons; first, it's specific to each database. Second, the number of change costs to be defined beforehand (this number is $C_2^n$, 2 should be above $n$ where $n$ is the size of the alphabet) can be somehow large. Third, if we try to use multiresolution techniques on the symbolic representation, then we will have to define a table for each resolution. Another serious problem arises in this case; merging two characters in text processing is not intuitional at all. So there's no clear way on how the "new" characters (those of a different resolution) can be related to the old ones.

In this paper, we present a new distance metric for symbolically represented data. It has a few advantages; one of them is dealing with the above problems in a natural way (no need to define a cost function for the change operation, no need to redefine it for different resolutions)

# 3   Background

Comparing strings of characters has several applications in computer science. These applications were in the beginning restricted to data structures whose representation is mainly symbolic (DNA and proteins sequences, textual data…etc). But later these applications were extended to other data structures that can be expressed symbolically.

There are a few distance metrics that deal with symbolically represented data. In this section, however, we will list the most common ones

## 3.1   The Longest Common Subsequence

Given two strings $S = [s_1, s_2, ..., s_m]$ and $R = [r_1, r_2, ..., r_n]$. Their longest common subsequence (abbreviated as LCSS) is the longest common subsequence to both of them. This subsequence doesn't have to be consecutive, but it has to have the same order in both strings.

It's important to notice that the LCSS uses the same operations that the ED uses, that's why many researchers consider the LCSS to be a special case of the ED

### 3.2 The Theory of Genome Rearrangements

Different species of organisms undergo a transformation process called evolution. An important problem in biology is to construct trees, called the phylogenetic trees, which help show how species are related to each other [14]. Different species are compared by comparing the sequences of their genomes.

Organisms evolve as a result of different evolutionary operations. An important operation is when a piece of the genome is reversed. During the long history of evolution this operation can happen several times. As a result to this, new species arise.

Genome sequences, because of their nature, are a data structure that can naturally be represented symbolically. There are many distance metrics used to compare two sequences of genomes (The exemplar distance, the breakpoint distance, signed reversal distance…etc). In general, these rearrangement distances measure the number of elementary operations necessary to transform one linear order on the genes into another [13]

## 4 The Proposed Distance

### 4.1 Introduction

We start by giving the following example;

Given the following string;

$$S_1 = marwan$$

By performing two change operations on $S_1$ in the first and fifth positions we obtain;

$$S_2 = aarwin$$

By calculating their edit distance we get; $ED(S_1, S_2) = 2$

Let $NC$ be the number of distinct characters that two (or more) strings contain, i.e.

$$NC = \left| \{ch(S_1)\} \cup \{ch(S_2)\} \right|$$

In our example we have;

$$NC(S_1, S_2) = 6$$

Now if we change the same positions in $S_1$ with different characters we obtain, for instance, the string

$S_3 = barwen$

By calculating the edit distance between $S_1$ and $S_3$ we get; $ED(S_1, S_3) = 2$ (which is the same as $ED(S_1, S_2)$ )

But we notice that

$NC(S_1, S_3) = 7\,7.$

This means that one change operation used a character that is more "familiar" to the two strings in the first case than in the second case, in other words, $S_2$ is closer to $S_1$ than $S_3$. However, the edit distance couldn't recognize this, since the edit distance was the same in both cases.

We will see later that this concept of "familiarity" can be extended to consider not only $NC$ but the frequency of sequences too.

N.B. We chose an example of strings of identical lengths since we were only discussing the change operation

### 4.2 Definition-The Extended Edit Distance

Let $A$ be a finite alphabet, and let $f_i^{(S)}$ be the frequency of the character $i$ in $S$, and $f_i^{(T)}$ be the frequency of the character $i$ in $T$, and where $S$, $T$ are two strings on $A$. The extended edit distance (EED) is defined as;

$$EED(S,T) = ED(S,T) + \lambda \left[ |S| + |T| - 2\sum_i \min(f_i^{(S)}, f_i^{(T)}) \right]$$

Where $|X|$ is the length of the string $X$, and where $\lambda \geq 0$ ($\lambda \in R$). We call $\lambda$ the frequency factor, and where $i$ is the number of elements of $NC$

*Remark :* Notice that when $\lambda = 0 \Rightarrow EED(S,T) = ED(S,T)$, and this is the minimum value for EED, so ED is actually a lower bound of EED

### 4.3 Theorem

The EED is a distance metric

**Proof:**

Before proving this theorem we notice that;

$$\lambda \left[ |S| + |T| - 2\sum_i \min(f_i^{(S)}, f_i^{(T)}) \right] \geq 0 \qquad \forall S, T \qquad \textbf{(1)}$$

(Obvious)

Now we start with the proof that EED is a distance metric.

i- $EED(S,T) = 0 \Leftrightarrow S = T$

a)  $EED(S,T) = 0 \Rightarrow S = T$

Proof:

If $EED(S,T) = 0$, and taking into account (1), then both (2) and (3) are valid;

$$|S| + |T| - 2\sum_i \min(f_i^{(S)}, f_i^{(T)}) = 0 \qquad \textbf{(2)}$$

$$ED(S,T) = 0 \qquad \textbf{(3)}$$

From (3), and since ED is a distance, we get;
$S = T$

b) $S = T \Rightarrow EED(S,T) = 0$
Proof:
Obvious

From (a) and (b) we have;

$EED(S,T) = 0 \Leftrightarrow S = T$

ii- $EED(S,T) = EED(T,S)$

Proof:
Obvious

iii- $EED(S,T) \leq EED(S,R) + EED(R,T)$

Proof:

$\forall S, T, R$ We have;

$$ED(S,T) \leq ED(S,R) + ED(R,T) \qquad \textbf{(4)}$$

(Valid since ED is a distance metric)

We also have ;

$$\lambda[|S| + |T| - 2\sum_i \min(f_i^{(S)}, f_i^{(T)})] \leq$$
$$\lambda[|S| + |R| - 2\sum_i \min(f_i^{(S)}, f_i^{(R)})] + \lambda[|R| + |T| - 2\sum_i \min(f_i^{(R)}, f_i^{(T)})] \qquad \textbf{(5)}$$

<span style="color:red">(See Appendix A for the proof of (5))</span>

Adding (4), (5) side to side we get;

$$EED(S,T) \leq EED(S,R) + EED(R,T)$$

From i,ii,iii, we conclude that theorem 4-3 holds

## Example 3

(Revisiting the example presented in section 4.1)

We define the form of a string is a vector as follows:

$Form(S) = [f_1, f_2, ..., f_n]$   ($n$ is the size of the alphabet, in our example it's 26, the English alphabet)

$Form(S_1) = [2, 0, ....., 0, \underset{M}{1}, \underset{N}{1}, 0, .., 0, \underset{R}{1}, 0, ....0, \underset{W}{1}, 0..]$

$Form(S_2) = [2, 0, ....., 0, 1, 0, .. 0, \underset{I}{1}, \underset{N}{1}, 0, .., 0, \underset{R}{1}, 0, ...., 0, \underset{w}{1}, 0...]$

$Form(S_3) = [\underset{A}{1}, \underset{B}{1}, 0, ...., 0, \underset{E}{1}, 0, ...., 0, \underset{N}{1}, 0, ..., 0, \underset{R}{1}, 0, ...., 0, \underset{W}{1}, 0....]$

( In this example,  $\lambda = 1$)
$EED(S_1, S_2) = 4$
$EED(S_1, S_3) = 6$

Which is what we're expecting, since, according to the concept of similarity we presented in section 4-1, $S_2$ is more similar to $S_1$ than $S_3$ is.

N.B. (in the example above we have $|S_1| = |S_2| = |S_3|$, but the EED can be applied to strings of different lengths, we chose this example because it was the example used in section 4-1)

**Example 4**

Given; $S_1 = marwan$, $S_2 = aarwin$ (The same $S_1$, $S_2$ as in the first example)
$S_3$ is obtained by changing $S_1$ in the same positions, but with different characters;

$S_3 = rarwen$
In this interesting example we have a particular case where ;
$NC(S_1, S_2) = NC(S_1, S_3) = 6$
But we still have

$EED(S_1, S_2) = 4$
$EED(S_1, S_3) = 6$
Which means that our proposed distance kept considering $S_2$ closer to $S_1$ than $S_3$ is, and this is what we expect since

$\left| \{char_a(S_1) \cap \{char_a(S_2)\} \right| = 2$ , $\left| \{char_a(S_1) \cap \{char_a(S_3)\} \right| = 1$ (you change the meaning of your notation sec. 4.1

So it's logical to consider $S_2$ closer to $S_1$ than $S_3$ is.

(For all other characters we have
$\left| \{char_x(S_1) \cap \{char_x(S_2)\} \right| = 1$ , $\left| \{char_x(S_1) \cap \{char_x(S_3)\} \right| = 1$ )

**Example 5**

We will give another example that will help show the properties of our EED;

Let
$S_1 = narwan$
$S_2 = aarwnn$
$S_3 = aarwxn$
$S_4 = xarwnn$
$S_5 = xarwxn$

The ED between $S_1$ and each of the other four strings is the same (which is 2). However, we will show that the EED is not the same, and that it differs according to how much each any two strings differ.

i- $EED(S_1, S_2) = 2$, which is the same as their ED. The two strings $S_1, S_2$ have the same length, the same characters, and the same frequency for each character (in fact, one string results from the other by rearranging it). These two strings have the highest hidden similarity of all the other pairs ($S_1$ and one of $S_2...S_5$) so their ED is their EED. This is the case that we mentioned in the remark in section 4-2

ii- $EED(S_1, S_3) = EDD(S_1, S_4) = 4$, each of $S_3, S_4$ results from $S_2$ by replacing one of the characters in $S_2$ by another character $x$ (in position 5 for $S_3$ and position 1 for $S_4$).

$x$ is a character that doesn't exist in $S_1$, so adding this "unfamiliar" character makes each of these strings less similar to $S_1$ than $S_2$ is. We also see from this case that the position at which this unfamiliar character was changed didn't affect the EED.

iii- If we continue this process and change the characters in position 4 in $S_4$ or in position 1 in $S_3$ with that same unfamiliar character $x$ (in both cases we obtain $S_5$). In both of these cases we substitute a familiar character ($a$ in the first case and $n$ in the second case) with an unfamiliar character $x$ so there should be loss of similarity compared with $S_3$ and $S_4$.

By calculating the EED we see that:

$EED(S_1, S_5) = 6$, which is what we expected.

We see that the EED was not the same in the above cases, while the ED was always the same.

**Example 6**

This example concerns strings of different lengths; let $S_1 = abca$, $S_2 = aabbcc$, $S_3 = adbecf$, we say that $ED(S_1, S_2) = ED(S_1, S_3) = 3$ However, by calculating their EED we find that $EED(S_1, S_2) = 5$ while $EED(S_1, S_3) = 7$

## 5  Complexity Analysis

The time complexity of EED is $O(m \times n)$, where $m$ is the length of the first time series and $n$ is the length of the second time series, or $O(n^2)$ if the two time series are of the same lengths. The complexity is high. However, we have to take into consideration that EED is a universal distance that can be applied to all symbolic represented data objects, where other distance measures are not applicable.

In order to make EED scale well when applied to time series, we can find a symbolic representation method that can allow high compression of the time series, with acceptable accuracy.

## 6  Experiments

We conducted two main experiments of times series classification task. As mentioned earlier, this new distance metric is applied to data structures which are represented symbolically, whether naturally or by using a certain technique of symbolic representation. We believe that bioinformatics or textual data bases are the ideal data structures to apply the EED to. However, and since our field of research is time series, we had to test EED on time series data bases.

Time series are not naturally represented symbolically. But more and more researchers are interested in symbolic representation of time series. A few methods have been proposed (see the introduction). Some of these methods are ad hoc, others are more sophisticated.

One of the most famous methods in the literature is SAX [7]. SAX, in simple words, consists of three steps;

1-Reducing the dimensionality of the time series by using PAA (After normalizing the times series)

2-Discretization the PAA to get a discrete representation of the times series(Using breakpoints)

3-Using a distance measure defined by the authors

To test EED we proceeded in the same way for steps 1 and 2 above to get a symbolic representation of time series, then in step 3 we compared EED with ED and the distance measure defined in SAX. On the resulting strings

In our experiments we tested our method on all the databases available at UCR Time Series Data Mining Archive [19] (except Plane and Car, since these two databases were not available).

The tests were aimed at comparing three main methods; the edit distance (ED) (we tested it for comparison reasons), our method; the extended edit distance (EED), and SAX . It's very important to point out that ED is mainly a method that is applied to textual data, what we did to test it on time series was to use the symbolic representation suggested in SAX, then we applied the ED to these symbolic representation obtained (the same thing we did to test EED). Anyway, SAX is a method that is designed directly to be used on time series, so it's a very competitive method.

**The First Experiment**

In order to make a fair comparison, we used the same compression ratio that was used to test SAX (i.e. 1:4) We also used the same range of alphabet size (3-10)

Both ED and SAX have one parameter which is the alphabet size, EED has one extra parameter, which is the frequency factor $\lambda$ (of course the three methods have another parameter which is the compression ratio)

For each dataset we test the parameters on the training set to get the optimal values of these parameters; the values that minimize the error. Then we use these optimal values on the testing set.

As for parameter $\lambda$ , for simplicity, we optimized it in the interval $[0,1]$ only, except in the cases where there was strong evidence that the error was decreasing monotonously as $\lambda$ increased

When comparing a method with another one there are two statistical parameters to be used, one of them is the mean error. The smaller the mean error is the better the method is. Another
statistical parameter is the standard deviation (STD). The importance of this latter is to show how universal the method is (i.e. can be applied to as many databases as possible). Here also, the smaller the STD is the better the method is.

After testing the parameters on the training sets of all the databases we got the following results (Table. 1) (There's no training for the Euclidean distance). The best method is highlighted

**Table 1**

|  | The Edit Distance (ED) | The Extended Edit Distance (EED) | SAX |
|---|---|---|---|
| Synthetic Control | 0.0367 | 0.0367 | 0.0267 |
| Gun-Point | 0.02 | 0.02 | 0.08 |
| CBF | 0.033 | 0 | 0.167 |
| Face (all) | 0.157 | 0.157 | 0.118 |
| OSULeaf | 0.2 | 0.195 | 0.365 |
| SwedishLeaf | 0.34 | 0.316 | 0.486 |
| 50words | 0.253 | 0.253 | 0.349 |
| Trace | 0.05 | 0.01 | 0.31 |
| Two_Patterns | 0.011 | 0.011 | 0.075 |
| Wafer | 0.002 | 0.002 | 0.005 |
| Face (four) | 0 | 0 | 0.208 |
| Lighting-2 | 0.15 | 0.117 | 0.267 |
| Lighting-7 | 0.4 | 0.371 | 0.371 |
| ECG200 | 0.15 | 0.13 | 0.14 |
| Adiac | 0.687 | 0.674 | 0.918 |
| Yoga | 0.193 | 0.193 | 0.24 |
| Fish | 0.194 | 0.194 | 0.509 |
| Beef | 0.533 | 0.433 | 0.467 |
| Coffee | 0 | 0 | 0.5 |
| OliveOil | 0.333 | 0.333 | 0.833 |
| MEAN | 0.195 | 0.181 | 0.322 |
| STD | 0.193 | 0.183 | 0.248 |

Then we used the optimal parameters for each dataset and for each method to apply them to the testing set and we got the following results (Table. 2);

**Table 2**

| | 1-NN Euclidean Distance | The Edit Distance (ED) | The Extended Edit Distance (EED) | SAX |
|---|---|---|---|---|
| Synthetic Control | 0.12 | 0.037 $\alpha^* = 7$ | 0.037 $\alpha = 7, \lambda = 0$ | 0.033 $\alpha = 10$ |
| Gun-Point | 0.087 | 0.073<br>$\alpha = 4$ | 0.06<br>$\alpha = 4, \lambda = 0.25$ | 0.233<br>$\alpha = 10$ |
| CBF | 0.148 | 0.029<br>$\alpha = 10$ | 0.026<br>$\alpha = 3, \lambda = 0.75$ | 0.104<br>$\alpha = 10$ |
| Face (all) | 0.286 | 0.324<br>$\alpha = 7$ | 0.324<br>$\alpha\ 7=, \lambda = 0$ | 0.319<br>$\alpha = 10$ |
| OSULeaf | 0.483 | 0.318<br>$\alpha = 5$ | 0.293<br>$\alpha = 5, \lambda = 0.75$ | 0.475<br>$\alpha = 9$ |
| SwedishLeaf | 0.213 | 0.344<br>$\alpha = 7$ | 0.365<br>$\alpha = 7, \lambda = 0.25$ | 0.490<br>$\alpha = 10$ |
| 50words | 0.369 | 0.266<br>$\alpha = 7$ | 0.266<br>$\alpha = 7, \lambda = 0$ | 0.327<br>$\alpha = 9$ |
| Trace | 0.24 | 0.11<br>$\alpha = 10$ | 0.07<br>$\alpha = 6, \lambda \geq 1.25$ | 0.42<br>$\alpha = 10$ |
| Two_Patterns | 0.09 | 0.015<br>$\alpha = 3$ | 0.015<br>$\alpha = 3, \lambda = 0$ | 0.081<br>$\alpha = 10$ |
| Wafer | 0.005 | 0.008<br>$\alpha = 4$ | 0.008<br>$\alpha = 4, \lambda = 0$ | 0.004<br>$\alpha = 6, 9$ |
| Face (four) | 0.216 | 0.045<br>$\alpha = 5$ | 0.045<br>$\alpha = 5, \lambda = 0,$ | 0.239<br>$\alpha = 3$ |
| Lighting-2 | 0.246 | 0.230<br>$\alpha = 10$ | 0.230<br>$\alpha = 7, \lambda = 1.75$ | 0.213<br>$\alpha = 6$ |
| Lighting-7 | 0.425 | 0.247<br>$\alpha = 10$ | 0.26<br>$\alpha = 4, \lambda = 0.75$ | 0.493<br>$\alpha = 7$ |
| ECG200 | 0.12 | 0.16<br>$\alpha = 6$ | 0.19<br>$\alpha = 5, \lambda = 0.25$ | 0.09<br>$\alpha = 10$ |
| Adiac | 0.389 | 0.701<br>$\alpha = 7$ | 0.642<br>$\alpha = 9, \lambda = 0.5$ | 0.903<br>$\alpha = 10$ |
| Yoga | 0.170 | 0.155<br>$\alpha = 7$ | 0.155<br>$\alpha = 7, \lambda = 0$ | 0.199<br>$\alpha = 10$ |
| Fish | 0.217 | 0.149<br>α =10 | 0.149<br>α =10, λ =0 | 0.514<br>α =10 |
| Beef | 0.467 | 0.467<br>α =4 | 0.4<br>α =4, λ =0.75 | 0.533<br>α =10 |
| Coffee | 0.25 | 0.107<br>α =8 | 0.107<br>α =8, λ =0 | 0.464<br>for all α |
| OliveOil | 0.133 | 0.467<br>α =9 | 0.4<br>α =9, λ ≥ 0.75 | 0.833<br>for all α |
| MEAN | 0.234 | 0.213 | 0.202 | 0.348 |
| STD | 0.134 | 0.183 | 0.168 | 0.247 |

\*: α is the alphabet size

The results obtained show that the average error is the smallest for EED, it's even smaller than that of the Euclidean distance. They also show that of all the three tested methods (ED,EED, and SAX) EED has the minimum standard deviation, which means that EED is the most universal one of the three tested methods.

It's worth mentioning for the Euclidian distance, there is no compression of information and in average it gives better results than symbolic compressed distances.

**The Second Experiment:**

In order to study the impact of using a different range of alphabet size, we chose, of the above datasets, those with a relatively large error. Our criterion was to choose the datasets were the error was large for at least two methods (this way we wouldn't be biasing any of the three methods). Our reference was the error of the Euclidean distance of that dataset. So, in order to see if the error will decrease when choosing a different alphabet range, we chose the datasets were the error was greater or equal to the error of the Euclidean distance for at least two of the three methods. So the datasets chosen are; FaceAll, SwedishLeaf, wafer, ECG200, Adiac, Beef, OliveOil (7 datasets)

It's important to mention here that even though the optimization process on the training set is actually a generalization of the optimization process of the first experiment (where the alphabet size was between 3 and 10), this second experiment is completely independent on the first one, since the parameters that optimize the training set of a certain dataset don't necessarily give the smallest error for the testing set. In fact, the error may even increase when using a wider range of alphabet size.

In order to study the impact of using a wider range of alphabet size, we calculate, on the train data, the mean and standard deviation of the error for the datasets in question, for an alphabet size varying in [3, 10 ] (Table. 3)

**Table 3**

|  | 1-NN Euclidean Distance | The Edit Distance (ED) | The Extended Edit Distance (EED) | SAX |
|---|---|---|---|---|
| Face (all) | 0.286 | 0.324 $\alpha =7$ | 0.324 $\alpha$ 7=, $\lambda =0$ | 0.319 $\alpha =10$ |
| SwedishLeaf | 0.213 | 0.344 $\alpha =7$ | 0.365 $\alpha =7$, $\lambda =0.25$ | 0.490 $\alpha =10$ |
| Wafer | 0.005 | 0.008 $\alpha =4$ | 0.008 $\alpha =4$, $\lambda =0$ | 0.004 $\alpha =6, 9$ |
| ECG200 | 0.12 | 0.16 $\alpha =6$ | 0.19 $\alpha =5$, $\lambda =0.25$ | 0.09 $\alpha =10$ |
| Adiac | 0.389 | 0.701 $\alpha =7$ | 0.642 $\alpha =9$, $\lambda =0.5$ | 0.903 $\alpha =10$ |
| Beef | 0.467 | 0.467 $\alpha =4$ | 0.4 $\alpha =4$, $\lambda =0.75$ | 0.533 $\alpha =10$ |
| OliveOil | 0.133 | 0.467 $\alpha =9$ | 0.4 $\alpha =9$, $\lambda \geq 0.75$ | 0.833 for all $\alpha$ |
| MEAN | 0.230 | 0.353 | 0.333 | 0.453 |
| STD | 0.162 | 0.225 | 0.197 | 0.343 |

Now, in order to study the error for the new range, we proceed in the same way we did for the first experiment, that is; we optimize the parameters on the training sets for the datasets in question, but this time for alphabet size that varies between 3 and 20, then we use these parameters on the testing sets of these databases, we get the following results (Table. 4)

**Table 4**

| | 1-NN Euclidean Distance | The Edit Distance (ED) | The Extended Edit Distance (EED) | SAX |
|---|---|---|---|---|
| Face (all) | 0.286 | 0.324 $\alpha$ =7 | 0.324 $\alpha$ =7, $\lambda$ =0 | 0.305 $\alpha$ =19 |
| SwedishLeaf | 0.213 | 0.344 $\alpha$ =7 | 0.365 $\alpha$ =7, $\lambda$ =0.25 | 0.253 $\alpha$ =20 |
| Wafer | 0.005 | 0.008 $\alpha$ =4 | 0.008 $\alpha$ =4, $\lambda$ =0 | 0.004 $\alpha$ =19 |
| ECG200 | 0.12 | 0.23 $\alpha$ =13 | 0.19 $\alpha$ =5, $\lambda$ =0.25 | 0.13 $\alpha$ =16 |
| Adiac | 0.389 | 0.555 $\alpha$ =18 | 0.524 $\alpha$ =19, $\lambda$ =1 | 0.867 $\alpha$ =18 |
| Beef | 0.467 | 0.467 $\alpha$ =17 | 0.4 $\alpha$ =4, $\lambda$ =0.75 | 0.433 $\alpha$ =20 |
| OliveOil | 0.133 | 0.333 $\alpha$ =16 | 0.333 $\alpha$ =16, $\lambda$=0, | 0.833 for all $\alpha$ |
| MEAN | 0.230 | 0.323 | 0.306 | 0.404 |
| STD | 0.162 | 0.175 | 0.165 | 0.333 |

This table shows that the average error has decreased when using an alphabet size between 3 and 20, so has the standard deviation, and for all the three methods. We also see that EED gave the smallest average error and standard deviation.

We got similar results when we tested some other datasets randomly. For example, when we tested Coffee on an alphabet size of [3, 20] we got error rate 0.071 for ED (alphabet size=12,13) , error rate 0 for EED (alphabet size=14, $\lambda$=0.25), and error rate 0.143 for SAX  (alphabet size=20)

We also tested other compression ratios on randomly chosen data sets (on alphabet size [3, 10]) , and we got similar results. Nevertheless, we didn't report these results, because these latter tests were not extensive.

## 7  Discussion

1- In the experiments we conducted we had to use time series of equal lengths for comparison reasons only, since SAX can be applied only to strings of equal lengths. But EED (and ED, too) can be applied to strings of different lengths.

2- It's worth mentioning that we didn't test alphabet size=2 because SAX is not applicable in this case (we can show that when alphabet size =2 then the distance between any two time series will be zero, and for any). Yet, we think it could be important for a method to be applicable for an alphabet size= 2, since this particular number is of interest in many applications.

3-We didn't report timing results, since the codes we used were not optimized for speed. But while testing the datasets, we noticed that SAX was faster than the other two methods.

4-Our method is not restricted to time series, and can be applied to other data types.

5- Our method, unlike SAX, is metric. However, the complexity of our method is $O(n^2)$ while that of SAX is $O(n)$

6-The main property of the EED over ED is that it is more precise, since it considers a global level of similarity by using the additional term:

$$|S| + |T| - 2\sum_i \min(f_i^{(S)}, f_i^{(T)})$$

This term decreases as the two strings have more and more general common features. This is very similar to the methods that use a similarity measure that takes into account not only the original data, but also different resolutions of it, all together represented by one vector.

7-With some datasets, SAX gave the very same error whatever the alphabet size was. Actually, that error was the same even when we tried a different compression ratio.

8-In order to represent the time series symbolically, we had to use a technique prepared for SAX, since this is the most famous symbolic representation technique of time series known in the literature. Nevertheless, a representation technique prepared mainly for EED may even give better results.

## 8  Future Work

The main advantage of the EED over the two other methods is that it can be extended to take into account not only the frequency of characters, but also the frequency of segments, so it can be applied to different resolutions, which is something we're working on.

Another possible future work is using the EED in anomaly detection in time series data mining, by representing the motif symbolically and applying the EED by taking the frequency of the motif rather than the frequency of characters

## 9  Conclusion

In this paper we presented a new distance metric applied to strings. The main feature of this distance is that it considers the frequency of characters, which is something other distance measures don't consider.

We tested this distance metric on a time series classification task, and we compared it to two other distances , and we showed that our distance gave better results, even when compared to a method (SAX) that is designed mainly for symbolically represented time series..

# Appendix A

Let $A$ be a finite alphabet, and let $f_i^{(S)}$ be the frequency of the character $i$ in $S$, where $S$ is a strings represented by $A$.

Let

$$D(S,T) = |S| + |T| - 2\sum_i \min(f_i^{(S)}, f_i^{(T)})$$

Then $\forall S_1, S_2, S_3$ we have;

$$D(S_1, S_2) \leq D(S_1, S_3) + D(S_3, S_2) \qquad \textbf{(1)}$$

for all $n$, where $n$ is the number of characters used to represent the strings

N.B. In the case when a character does not exist in one or more of the strings, we can assume that this (these) string(s) has (have) 0 frequency of the missing character.

**Proof**

The will prove the above lemma by induction.

i- $n = 1$

This is a trivial case. Given three strings; $\forall S_1, S_2, S_3$ represented by the same character $a$

Let $S_1^a, S_2^a, S_3^a$ be the frequency of $a$ in $\forall S_1, S_2, S_3$, respectively.

We have six configurations of this case;

1- $S_1^a \leq S_2^a \leq S_3^a$

2- $S_1^a \leq S_3^a \leq S_2^a$

3- $S_2^a \leq S_1^a \leq S_3^a$

4- $S_2^a \leq S_3^a \leq S_1^a$

5- $S_3^a \leq S_1^a \leq S_2^a$

6- $S_3^a \leq S_2^a \leq S_1^a$

We will prove that relation (1) holds in these six configurations.

1- $S_1^a \leq S_2^a \leq S_3^a$

In this case we have;

$\min(S_1^a, S_2^a) = S_1^a$

$\min(S_1^a, S_3^a) = S_1^a$

$\min(S_2^a, S_3^a) = S_2^a$

$D(S_1, S_2) \overset{?}{\leq} D(S_1, S_3) + D(S_3, S_2)$

By substituting the above values in this last relation we get;

$S_1^a + S_2^a - 2S_1^a \overset{?}{\leq} S_1^a + S_3^a - 2S_1^a + S_3^a + S_2^a - 2S_2^a$

$0 \overset{?}{\leq} 2S_3^a - 2S_2^a$

This is valid according to the stipulation of this configuration.

2- $S_1^a \leq S_3^a \leq S_2^a$

In this case we have;

$\min(S_1^a, S_2^a) = S_1^a$

$\min(S_1^a, S_3^a) = S_1^a$

$\min(S_2^a, S_3^a) = S_3^a$

$D(S_1, S_2) \overset{?}{\leq} D(S_1, S_3) + D(S_3, S_2)$

By substituting the above values in this last relation we get;

$S_1^a + S_2^a - 2S_1^a \overset{?}{\leq} S_1^a + S_3^a - 2S_1^a + S_3^a + S_2^a - 2S_3^a$

$0 \overset{?}{\leq} 0$

valid

3- $S_2^a \leq S_1^a \leq S_3^a$

In this case we have;

$\min(S_1^a, S_2^a) = S_2^a$

$\min(S_1^a, S_3^a) = S_1^a$

$\min(S_2^a, S_3^a) = S_2^a$

$D(S_1, S_2) \overset{?}{\leq} D(S_1, S_3) + D(S_3, S_2)$

By substituting the above values in this last relation we get;

$S_1^a + S_2^a - 2S_2^a \overset{?}{\leq} S_1^a + S_3^a - 2S_1^a + S_3^a + S_2^a - 2S_2^a$

$0 \overset{?}{\leq} 2S_3^a - 2S_1^a$

valid

4- $S_2^a \leq S_3^a \leq S_1^a$

In this case we have;

$\min(S_1^a, S_2^a) = S_2^a$

$\min(S_1^a, S_3^a) = S_3^a$

$\min(S_2^a, S_3^a) = S_2^a$

$D(S_1, S_2) \overset{?}{\leq} D(S_1, S_3) + D(S_3, S_2)$

By substituting the above values in this last relation we get;

$S_1^a + S_2^a - 2S_2^a \overset{?}{\leq} S_1^a + S_3^a - 2S_3^a + S_3^a + S_2^a - 2S_2^a$

$0 \overset{?}{\leq} 0$

valid

5- $S_3^a \leq S_1^a \leq S_2^a$

In this case we have;

$\min(S_1^a, S_2^a) = S_1^a$

$\min(S_1^a, S_3^a) = S_3^a$

$\min(S_2^a, S_3^a) = S_3^a$

$D(S_1, S_2) \overset{?}{\leq} D(S_1, S_3) + D(S_3, S_2)$

By substituting the above values in this last relation we get;

$S_1^a + S_2^a - 2S_1^a \overset{?}{\leq} S_1^a + S_3^a - 2S_3^a + S_3^a + S_2^a - 2S_3^a$

$0 \overset{?}{\leq} 2S_1^a - 2S_3^a$

valid

6- $S_3^a \leq S_2^a \leq S_1^a$

In this case we have;

$\min(S_1^a, S_2^a) = S_2^a$

$\min(S_1^a, S_3^a) = S_3^a$

$\min(S_2^a, S_3^a) = S_3^a$

$D(S_1, S_2) \overset{?}{\leq} D(S_1, S_3) + D(S_3, S_2)$

By substituting the above values in this last relation we get;

$S_1^a + S_2^a - 2S_2^a \overset{?}{\leq} S_1^a + S_3^a - 2S_3^a + S_3^a + S_2^a - 2S_3^a$

$0 \overset{?}{\leq} 2S_2^a - 2S_3^a$

valid

From 1-6 we conclude that the lemma is valid for $n = 1$

ii- Let's assume that the lemma holds for $n - 1$, where $n \geq 2$ and we will prove it for $n$

Since the lemma holds for $n - 1$ then ;

$$D(S_1, S_2) \leq D(S_1, S_3) + D(S_3, S_2) \qquad \textbf{(2)}$$

where

$$D(S_1, S_2) = |S_1| + |S_2| - 2 \sum_{i=1}^{n-1} \min(f_i^{(S_1)}, f_i^{(S_2)})$$

$$D(S_1, S_3) = |S_1| + |S_3| - 2 \sum_{i=1}^{n-1} \min(f_i^{(S_1)}, f_i^{(S_3)})$$

$$D(S_3, S_2) = |S_3| + |S_2| - 2 \sum_{i=1}^{n-1} \min(f_i^{(S_3)}, f_i^{(S_2)})$$

When a new character is added the strings represented by $n-1$ characters are represented by $n$

Let the frequency of the newly introduced character be $f_n^{(S_1)}, f_n^{(S_2)}, f_n^{(S_3)}$ in $\forall S_1, S_2, S_3$ respectively.

We have six configurations of the newly added character;

7- $f_n^{(S_1)} \leq f_n^{(S_2)} \leq f_n^{(S_3)}$

8- $f_n^{(S_1)} \leq f_n^{(S_3)} \leq f_n^{(S_2)}$

9- $f_n^{(S_2)} \leq f_n^{(S_1)} \leq f_n^{(S_3)}$

10- $f_n^{(S_2)} \leq f_n^{(S_3)} \leq f_n^{(S_1)}$

11- $f_n^{(S_3)} \leq f_n^{(S_1)} \leq f_n^{(S_2)}$

12- $f_n^{(S_3)} \leq f_n^{(S_2)} \leq f_n^{(S_1)}$

We will prove that relation (1) holds in these six configurations.

7- $f_n^{(S_1)} \leq f_n^{(S_2)} \leq f_n^{(S_3)}$

In this case we have;

$$\min(f_n^{(S_1)}, f_n^{(S_2)}) = f_n^{(S_1)}$$

$$\min(f_n^{(S_1)}, f_n^{(S_3)}) = f_n^{(S_1)}$$

$$\min(f_n^{(S_2)}, f_n^{(S_3)}) = f_n^{(S_2)}$$

$$D(S_1,S_2) \overset{?}{\leq} D(S_1,S_3) + D(S_3,S_2)$$
$$\Rightarrow$$

$$|S_1|+|S_2|-2\sum_{i=1}^{n-1}\min(f_i^{(S_1)},f_i^{(S_2)})+f_n^{(S_1)}+f_n^{(S_2)}-2\min(f_n^{(S_1)},f_n^{(S_2)}) \overset{?}{\leq}$$

$$|S_1|+|S_3|-2\sum_{i=1}^{n-1}\min(f_i^{(S_1)},f_i^{(S_3)})+f_n^{(S_1)}+f_n^{(S_3)}-2\min(f_n^{(S_1)},f_n^{(S_3)})+$$

$$|S_3|+|S_2|-2\sum_{i=1}^{n-1}\min(f_i^{(S_3)},f_i^{(S_2)})+f_n^{(S_3)}+f_n^{(S_2)}-2\min(f_n^{(S_3)},f_n^{(S_2)})$$
$$\Longrightarrow$$

$$|S_1|+|S_2|-2\sum_{i=1}^{n-1}\min(f_i^{(S_1)},f_i^{(S_2)})+f_n^{(S_1)}+f_n^{(S_2)}-2f_n^{(S_1)} \overset{?}{\leq}$$

$$|S_1|+|S_3|-2\sum_{i=1}^{n-1}\min(f_i^{(S_1)},f_i^{(S_3)})+f_n^{(S_1)}+f_n^{(S_3)}-2f_n^{(S_1)}$$

$$|S_3|+|S_2|-2\sum_{i=1}^{n-1}\min(f_i^{(S_3)},f_i^{(S_2)})+f_n^{(S_3)}+f_n^{(S_2)}-2f_n^{(S_2)}$$

$$\Longrightarrow$$

$$|S_1|+|S_2|-2\sum_{i=1}^{n-1}\min(f_i^{(S_1)},f_i^{(S_2)}) \overset{?}{\leq}$$

$$|S_1|+|S_3|-2\sum_{i=1}^{n-1}\min(f_i^{(S_1)},f_i^{(S_3)})$$

$$|S_3|+|S_2|-2\sum_{i=1}^{n-1}\min(f_i^{(S_3)},f_i^{(S_2)})+2f_n^{(S_3)}-2f_n^{(S_2)}$$

Taking (2) into account , we get;

$$0 \overset{?}{\leq} 2f_n^{(S_3)}-2f_n^{(S_2)}$$

which is valid according to (7)

8- $f_n^{(S_1)} \leq f_n^{(S_3)} \leq f_n^{(S_2)}$

In this case we have

$$\min(f_n^{(S_1)},f_n^{(S_2)}) = f_n^{(S_1)}$$

$$\min(f_n^{(S_1)}, f_n^{(S_3)}) = f_n^{(S_1)}$$

$$\min(f_n^{(S_2)}, f_n^{(S_3)}) = f_n^{(S_3)}$$

$$D(S_1, S_2) \overset{?}{\leq} D(S_1, S_3) + D(S_3, S_2)$$
$$\Rightarrow$$

$$|S_1| + |S_2| - 2\sum_{i=1}^{n-1}\min(f_i^{(S_1)}, f_i^{(S_2)}) + f_n^{(S_1)} + f_n^{(S_2)} - 2\min(f_n^{(S_1)}, f_n^{(S_2)}) \overset{?}{\leq}$$

$$|S_1| + |S_3| - 2\sum_{i=1}^{n-1}\min(f_i^{(S_1)}, f_i^{(S_3)}) + f_n^{(S_1)} + f_n^{(S_3)} - 2\min(f_n^{(S_1)}, f_n^{(S_3)}) +$$

$$|S_3| + |S_2| - 2\sum_{i=1}^{n-1}\min(f_i^{(S_3)}, f_i^{(S_2)}) + f_n^{(S_3)} + f_n^{(S_2)} - 2\min(f_n^{(S_3)}, f_n^{(S_2)})$$
$$\Rightarrow$$

$$|S_1| + |S_2| - 2\sum_{i=1}^{n-1}\min(f_i^{(S_1)}, f_i^{(S_2)}) + f_n^{(S_1)} + f_n^{(S_2)} - 2f_n^{(S_1)} \overset{?}{\leq}$$

$$|S_1| + |S_3| - 2\sum_{i=1}^{n-1}\min(f_i^{(S_1)}, f_i^{(S_3)}) + f_n^{(S_1)} + f_n^{(S_3)} - 2f_n^{(S_1)}$$

$$|S_3| + |S_2| - 2\sum_{i=1}^{n-1}\min(f_i^{(S_3)}, f_i^{(S_2)}) + f_n^{(S_3)} + f_n^{(S_2)} - 2f_n^{(S_3)}$$

$$\Rightarrow$$

$$|S_1| + |S_2| - 2\sum_{i=1}^{n-1}\min(f_i^{(S_1)}, f_i^{(S_2)}) \overset{?}{\leq}$$

$$|S_1| + |S_3| - 2\sum_{i=1}^{n-1}\min(f_i^{(S_1)}, f_i^{(S_3)})$$

$$|S_3| + |S_2| - 2\sum_{i=1}^{n-1}\min(f_i^{(S_3)}, f_i^{(S_2)})$$

Which is true according to (2)

9- $f_n^{(S_2)} \leq f_n^{(S_1)} \leq f_n^{(S_3)}$

In this case we have

$$\min(f_n^{(S_1)}, f_n^{(S_2)}) = f_n^{(S_2)}$$

$$\min(f_n^{(S_1)}, f_n^{(S_3)}) = f_n^{(S_1)}$$

$\min(f_n^{(S_2)}, f_n^{(S_3)}) = f_n^{(S_2)}$

$D(S_1, S_2) \overset{?}{\leq} D(S_1, S_3) + D(S_3, S_2)$
$\Rightarrow$

$|S_1| + |S_2| - 2\sum_{i=1}^{n-1}\min(f_i^{(S_1)}, f_i^{(S_2)}) + f_n^{(S_1)} + f_n^{(S_2)} - 2\min(f_n^{(S_1)}, f_n^{(S_2)}) \overset{?}{\leq}$

$|S_1| + |S_3| - 2\sum_{i=1}^{n-1}\min(f_i^{(S_1)}, f_i^{(S_3)}) + f_n^{(S_1)} + f_n^{(S_3)} - 2\min(f_n^{(S_1)}, f_n^{(S_3)}) +$

$|S_3| + |S_2| - 2\sum_{i=1}^{n-1}\min(f_i^{(S_3)}, f_i^{(S_2)}) + f_n^{(S_3)} + f_n^{(S_2)} - 2\min(f_n^{(S_3)}, f_n^{(S_2)})$
$\Rightarrow$

$|S_1| + |S_2| - 2\sum_{i=1}^{n-1}\min(f_i^{(S_1)}, f_i^{(S_2)}) + f_n^{(S_1)} + f_n^{(S_2)} - 2f_n^{(S_2)} \overset{?}{\leq}$

$|S_1| + |S_3| - 2\sum_{i=1}^{n-1}\min(f_i^{(S_1)}, f_i^{(S_3)}) + f_n^{(S_1)} + f_n^{(S_3)} - 2f_n^{(S_1)}$

$|S_3| + |S_2| - 2\sum_{i=1}^{n-1}\min(f_i^{(S_3)}, f_i^{(S_2)}) + f_n^{(S_3)} + f_n^{(S_2)} - 2f_n^{(S_2)}$

$\Rightarrow$

$|S_1| + |S_2| - 2\sum_{i=1}^{n-1}\min(f_i^{(S_1)}, f_i^{(S_2)}) \overset{?}{\leq}$

$|S_1| + |S_3| - 2\sum_{i=1}^{n-1}\min(f_i^{(S_1)}, f_i^{(S_3)})$

$|S_3| + |S_2| - 2\sum_{i=1}^{n-1}\min(f_i^{(S_3)}, f_i^{(S_2)}) + 2f_n^{(S_3)} - 2f_n^{(S_1)}$

Taking (2) into account , we get;

$0 \overset{?}{\leq} 2f_n^{(S_3)} - 2f_n^{(S_1)}$

which is valid according to (9)

10- $f_n^{(S_2)} \leq f_n^{(S_3)} \leq f_n^{(S_1)}$

In this case we have

$\min(f_n^{(S_1)}, f_n^{(S_2)}) = f_n^{(S_2)}$

$$\min(f_n^{(S_1)}, f_n^{(S_3)}) = f_n^{(S_3)}$$

$$\min(f_n^{(S_2)}, f_n^{(S_3)}) = f_n^{(S_2)}$$

$$D(S_1, S_2) \overset{?}{\leq} D(S_1, S_3) + D(S_3, S_2)$$
$$\Rightarrow$$

$$|S_1| + |S_2| - 2\sum_{i=1}^{n-1}\min(f_i^{(S_1)}, f_i^{(S_2)}) + f_n^{(S_1)} + f_n^{(S_2)} - 2\min(f_n^{(S_1)}, f_n^{(S_2)}) \overset{?}{\leq}$$

$$|S_1| + |S_3| - 2\sum_{i=1}^{n-1}\min(f_i^{(S_1)}, f_i^{(S_3)}) + f_n^{(S_1)} + f_n^{(S_3)} - 2\min(f_n^{(S_1)}, f_n^{(S_3)}) +$$

$$|S_3| + |S_2| - 2\sum_{i=1}^{n-1}\min(f_i^{(S_3)}, f_i^{(S_2)}) + f_n^{(S_3)} + f_n^{(S_2)} - 2\min(f_n^{(S_3)}, f_n^{(S_2)})$$
$$\Rightarrow$$

$$|S_1| + |S_2| - 2\sum_{i=1}^{n-1}\min(f_i^{(S_1)}, f_i^{(S_2)}) + f_n^{(S_1)} + f_n^{(S_2)} - 2f_n^{(S_2)} \overset{?}{\leq}$$

$$|S_1| + |S_3| - 2\sum_{i=1}^{n-1}\min(f_i^{(S_1)}, f_i^{(S_3)}) + f_n^{(S_1)} + f_n^{(S_3)} - 2f_n^{(S_3)}$$

$$|S_3| + |S_2| - 2\sum_{i=1}^{n-1}\min(f_i^{(S_3)}, f_i^{(S_2)}) + f_n^{(S_3)} + f_n^{(S_2)} - 2f_n^{(S_2)}$$
$$\Rightarrow$$

$$|S_1| + |S_2| - 2\sum_{i=1}^{n-1}\min(f_i^{(S_1)}, f_i^{(S_2)}) \overset{?}{\leq}$$

$$|S_1| + |S_3| - 2\sum_{i=1}^{n-1}\min(f_i^{(S_1)}, f_i^{(S_3)})$$

$$|S_3| + |S_2| - 2\sum_{i=1}^{n-1}\min(f_i^{(S_3)}, f_i^{(S_2)})$$

Which is true according to (2)

11- $f_n^{(S_3)} \leq f_n^{(S_1)} \leq f_n^{(S_2)}$

In this case we have

$$\min(f_n^{(S_1)}, f_n^{(S_2)}) = f_n^{(S_1)}$$

$$\min(f_n^{(S_1)}, f_n^{(S_3)}) = f_n^{(S_3)}$$

$$\min(f_n^{(S_2)}, f_n^{(S_3)}) = f_n^{(S_3)}$$

$$D(S_1,S_2) \overset{?}{\leq} D(S_1,S_3) + D(S_3,S_2)$$
$$\Rightarrow$$

$$|S_1| + |S_2| - 2\sum_{i=1}^{n-1}\min(f_i^{(S_1)}, f_i^{(S_2)}) + f_n^{(S_1)} + f_n^{(S_2)} - 2\min(f_n^{(S_1)}, f_n^{(S_2)}) \overset{?}{\leq}$$

$$|S_1| + |S_3| - 2\sum_{i=1}^{n-1}\min(f_i^{(S_1)}, f_i^{(S_3)}) + f_n^{(S_1)} + f_n^{(S_3)} - 2\min(f_n^{(S_1)}, f_n^{(S_3)}) +$$

$$|S_3| + |S_2| - 2\sum_{i=1}^{n-1}\min(f_i^{(S_3)}, f_i^{(S_2)}) + f_n^{(S_3)} + f_n^{(S_2)} - 2\min(f_n^{(S_3)}, f_n^{(S_2)})$$

$$\Longrightarrow$$

$$|S_1| + |S_2| - 2\sum_{i=1}^{n-1}\min(f_i^{(S_1)}, f_i^{(S_2)}) + f_n^{(S_1)} + f_n^{(S_2)} - 2f_n^{(S_1)} \overset{?}{\leq}$$

$$|S_1| + |S_3| - 2\sum_{i=1}^{n-1}\min(f_i^{(S_1)}, f_i^{(S_3)}) + f_n^{(S_1)} + f_n^{(S_3)} - 2f_n^{(S_3)}$$

$$|S_3| + |S_2| - 2\sum_{i=1}^{n-1}\min(f_i^{(S_3)}, f_i^{(S_2)}) + f_n^{(S_3)} + f_n^{(S_2)} - 2f_n^{(S_3)}$$
$$\Longrightarrow$$

$$|S_1| + |S_2| - 2\sum_{i=1}^{n-1}\min(f_i^{(S_1)}, f_i^{(S_2)}) \overset{?}{\leq}$$

$$|S_1| + |S_3| - 2\sum_{i=1}^{n-1}\min(f_i^{(S_1)}, f_i^{(S_3)})$$

$$|S_3| + |S_2| - 2\sum_{i=1}^{n-1}\min(f_i^{(S_3)}, f_i^{(S_2)}) + 2f_n^{(S_1)} - 2f_n^{(S_3)}$$

Taking (2) into account , we get;

$$0 \overset{?}{\leq} 2f_n^{(S_1)} - 2f_n^{(S_3)}$$

valid according to (11)

12- $f_n^{(S_3)} \leq f_n^{(S_2)} \leq f_n^{(S_1)}$

In this case we have

$$\min(f_n^{(S_1)}, f_n^{(S_2)}) = f_n^{(S_2)}$$

$$\min(f_n^{(S_1)}, f_n^{(S_3)}) = f_n^{(S_3)}$$

$$\min(f_n^{(S_2)}, f_n^{(S_3)}) = f_n^{(S_3)}$$

$$D(S_1, S_2) \overset{?}{\leq} D(S_1, S_3) + D(S_3, S_2)$$

$$|S_1| + |S_2| - 2\sum_{i=1}^{n-1} \min(f_i^{(S_1)}, f_i^{(S_2)}) + f_n^{(S_1)} + f_n^{(S_2)} - 2\min(f_n^{(S_1)}, f_n^{(S_2)}) \overset{?}{\leq}$$

$$|S_1| + |S_3| - 2\sum_{i=1}^{n-1} \min(f_i^{(S_1)}, f_i^{(S_3)}) + f_n^{(S_1)} + f_n^{(S_3)} - 2\min(f_n^{(S_1)}, f_n^{(S_3)}) +$$

$$|S_3| + |S_2| - 2\sum_{i=1}^{n-1} \min(f_i^{(S_3)}, f_i^{(S_2)}) + f_n^{(S_3)} + f_n^{(S_2)} - 2\min(f_n^{(S_3)}, f_n^{(S_2)})$$

$$\Longrightarrow$$

$$|S_1| + |S_2| - 2\sum_{i=1}^{n-1} \min(f_i^{(S_1)}, f_i^{(S_2)}) + f_n^{(S_1)} + f_n^{(S_2)} - 2f_n^{(S_2)} \overset{?}{\leq}$$

$$|S_1| + |S_3| - 2\sum_{i=1}^{n-1} \min(f_i^{(S_1)}, f_i^{(S_3)}) + f_n^{(S_1)} + f_n^{(S_3)} - 2f_n^{(S_3)}$$

$$|S_3| + |S_2| - 2\sum_{i=1}^{n-1} \min(f_i^{(S_3)}, f_i^{(S_2)}) + f_n^{(S_3)} + f_n^{(S_2)} - 2f_n^{(S_3)}$$

$$\Longrightarrow$$

$$|S_1| + |S_2| - 2\sum_{i=1}^{n-1} \min(f_i^{(S_1)}, f_i^{(S_2)}) \overset{?}{\leq}$$

$$|S_1| + |S_3| - 2\sum_{i=1}^{n-1} \min(f_i^{(S_1)}, f_i^{(S_3)})$$

$$|S_3| + |S_2| - 2\sum_{i=1}^{n-1} \min(f_i^{(S_3)}, f_i^{(S_2)}) + 2f_n^{(S_1)} - 2f_n^{(S_3)}$$

Taking (2) into account , we get;

$$0 \overset{?}{\leq} 2f_n^{(S_2)} - 2f_n^{(S_3)}$$

which is true according to (12)

From 7-12 we conclude that the lemma is valid for $n$

From i and ii, the lemma holds